\documentclass[5p,twocolumn]{elsarticle}

\usepackage{graphicx}
\usepackage{dcolumn}
\usepackage{pifont}
\usepackage{bm}
\usepackage{multirow}
\usepackage{amsmath}
\usepackage{float}
\usepackage{txfonts}
\usepackage[version=3]{mhchem}

\journal{Inorganic Chemistry}

\begin{document}
\title{Two-dimensional hybrid composites of \ce{SnS2} with graphene and graphene oxide for improving sodium storage: A first-principles study}

\author[kimuniv]{Kum-Chol Ri}
\author[kimuniv]{Chol-Jun Yu\corref{cor}}
\ead{cj.yu@ryongnamsan.edu.kp}
\author[kimuniv]{Jin-Song Kim}
\author[kimuniv]{Song-Hyok Choe}

\cortext[cor]{Corresponding author}

\address[kimuniv]{Chair of Computational Materials Design, Faculty of Materials Science, Kim Il Sung University, Ryongnam-Dong, Taesong District, Pyongyang, Democratic People's Republic of Korea}

\begin{abstract}
Among the recent achievements of sodium-ion battery (SIB) electrode materials, hybridization of two-dimentional (2D) materials is one of the most interesting appointments.
In this work, we propose to use the 2D hybrid composites of \ce{SnS2} with graphene or graphene oxide (GO) layers as SIB anode, based on the first-principles calculations of their atomic structures, sodium intercalation energetics and electronic properties.
The calculations reveal that graphene or GO film can effectively support not only the stable formation of hetero-interface with the \ce{SnS2} layer but also the easy intercalation of sodium atom with low migration energy and acceptable low volume change.
The electronic charge density differences and the local density of state indicate that the electrons are transferred from the graphene or GO layer to the \ce{SnS2} layer, facilitating the formation of hetero-interface and improving the electronic conductance of the semiconducting \ce{SnS2} layer.
These 2D hybrid composites of \ce{SnS2}/G or GO are concluded to be more promising candidates for SIB anodes compared with the individual monolayers.
\end{abstract}

\begin{keyword}
Sodium-ion battery \sep Anode \sep Tin oxide \sep Graphene oxide \sep First-principles method
\end{keyword}

\maketitle

\section{Introduction}
Recently, inorganic two-dimentional (2D) materials have attracted remarkable attention for various energy applications including energy harvesting and energy storage devices due to their unique structural and electrochemical properties~\cite{Shi17jmca,Kang17jmca,Georgakilas16cr,Gursel15dm}.
Graphene and chemically modified graphene such as graphene oxide (GO) or reduced graphene oxide (rGO) are regarded as the representative 2D materials, being characterized by layer-stacked structures coupled by weak interlayer van der Waals (vdW) and strong in-plane covalent bonding interactions~\cite{Yu17nrm,Kaplan17csr,Ambrosi16csr,Zhu13jpcc,Dabhi14jap}.
Such graphene-like 2D structures can be found in elemental analogues of graphene (silicene, germanene, phosphorene, borophene), transition metal oxides (TMOs)~\cite{Kubota14mb}, transition metal dichalcogenides (TMDs)~\cite{Chhowalla13nm,Kang17jmca,Pumera14jmca}, and transition metal carbides/nitrides (MXenes)~\cite{Naguib14am}.
These 2D materials can be obtained by exfoliation or synthesis, and can provide large surface area with abundant anchoring sites and low volume change during the de/intercalation processes of atoms or molecules.
Therefore, 2D materials can be used as potential electrode materials for alkali-ion batteries with remarkably large capacity, high rate capability, good cycle performance and low price.

Many of 2D materials such as rGO~\cite{Ali17sr,TLiu14jpcl,YXWang13c}, layered vanadium pentoxide (\ce{V2O5})~\cite{Feng11jacs} and molybdenum disulfide (\ce{MoS2})~\cite{Mortazavi14jps} have been verified to be used as efficient electrode materials for lithium-ion batteries (LIBs) and sodium-ion batteries (SIBs) by experimental and theoretical studies.
In fact, as the most widely used and standard anode material for LIBs, graphite has the theoretical capacity of 372 mA$\cdot$h$\cdot$g$^{-1}$, which is not high enough to satisfy the growing demand of battery capacity.
For SIBs, moreover, it is almost impossible to intercalate sodium itself into graphite due to a big mismatch between the ionic radius of \ce{Na+} cation and the interlayer distance of graphite, and thus co-intercalation with organic molecules has been devised~\cite{yucj06,yucj14,yucj20}.
Alternatively, graphene and its derivatives such as GO or rGO, which can be obtained by exfoliation from graphite, have been studied extensively, confirming the doubled capacity compared with graphite, very high surface area (2680 m$^2\cdot$g$^{-1}$), high electron conductivity and excellent chemical stability~\cite{Ali17sr,Lin16nn,Bonaccorso15s,TLiu14jpcl,YXWang13c}.
However, both Li and Na bind very weakly to the pristine graphene or GO, resulting in Li/Na clustering and dendrite growth.
On the other hand, tin disulfide (\ce{SnS2}) as the typical TMD is recognized as a promising anode material due to its high initial capacity with an unique mechanism of alkali cation storage reaction~\cite{Kim07jps,Jung08am,Zhong12jpcc}.
The bulk \ce{SnS2} has \ce{PbI2}-type layered structure and exhibits a high lithium storage capacity (average capacity 583 mA$\cdot$h$\cdot$g$^{-1}$ after 30th cycles) and a good cycling stability (cycle life performance 85\% after 30 cycles)~\cite{Jung08am}, being much better performance than the conventional graphite electrodes.
The lithium or sodium storage in \ce{SnS2} is performed through the two-step reactions; (1) conversion of \ce{SnS2} into metallic Sn and (2) alloying with lithium or sodium forming Li-Sn or Na-Sn alloy.
By contrast, only the conversion reaction occurs during the lithium or sodium storage  in other TMDs such as \ce{MoS2}, \ce{WS2} and \ce{VS2}~\cite{Pumera14jmca}.
The reversible alloying reaction was found to give a large capacity of sodium storage in the first cycle, but to be accompanied with a large volume expansion (e.g., 420\% for the case of Na-Sn alloying~\cite{Jiang12nl,Ellis12jes}), leading to a poor capacity after a few cycles.

Designing hetero-layered architectures by coupling graphene or GO with TMD or TMO 2D materials has emerged as an effective strategy to enhance the Li/Na binding strength and reduce the volume change as well~\cite{Pomerantseva17ne,Himani16ao,Pan14ic,Zhou14ic}.
In particular, the hybrid composites formed by coupling rGO and TMDs have been reported to show a good performance as anode materials for LIBs and SIBs~\cite{Geim13n,David14an,Xie15afm}.
In such composite systems, graphene-based 2D materials can not only serve as a matrix to anchor the TMD layers but also improve the electrochemical performance of semiconducting TMDs.
As a typical example of such composites, therefore, the hybrid \ce{SnS2}/rGO composites including the hetero-interface can be constructed and used as excellent anode materials by combining the metallic electrical conductivity of graphene with the high storage capability of \ce{SnS2}~\cite{Chang11jps,Ma15cm,Qu14am,Zhang15afm,Xu16jac,Luo12ees}.
These synthesized \ce{SnS2}/rGO hybrids have shown to exhibit the excellent electrochemical performance such as high sodium storage capacity of 649 mA$\cdot$h$\cdot$g$^{-1}$ at the current density of 100 mA$\cdot$g$^{-1}$, ultra-long cycle life of 89\% and 69\% after 400 and 1000 cycles, and superior rate capability of 337 mA$\cdot$h$\cdot$g$^{-1}$ at the higher current density of 12.8 A$\cdot$g$^{-1}$ (28C) in only 1.3 minutes~\cite{Ma15cm,Qu14am,Zhang15afm,Xu16jac}.
To the best of our knowledge, no theoretical works for 2D hybrid \ce{SnS2}/G or GO composites can be found in literature in spite of such successful experiments and the great importance of clarifying the operational mechanism, motivating us to investigate these hetero-interfaces in atomic scale by means of first-principles modeling and simulations.

In this work, we make a modeling of hybrid \ce{SnS2}/G and \ce{SnS2}/GO composites, together with 2D bilayered \ce{SnS2}, for conducting comparative and systematic study by using first-principles calculations within the density functional theory (DFT).
Performing the structural optimizations, we analyze the detailed atomic structures of these 2D layered compounds, such as interlayer distance and bond lengths.
The activation barriers for the in-plane atomic migrations of sodium atom are calculated in these composite systems of \ce{SnS2}/G and \ce{SnS2}/GO, and in the \ce{SnS2} bilayer.
The local density of states (LDOS) and electronic charge desity differences are analyzed.
Based on the calculation data, we reveal the crucial role of hetero-interface behind the Na-storage enhancement of \ce{SnS2}/GO hybrid.

\section{Computational Methods}
Regarding the atomistic modeling, three kinds of 2D systems, {\it i.e.}, \ce{SnS2}, \ce{SnS2}/G and \ce{SnS2}/GO, were built using the corresponding supercells.
The \ce{SnS2} crystal has the hexagonal structure ($P\bar{3}m1$ space group) with the lattice constants of $a=3.649$ \AA~and $c=5.899$ \AA~\cite{Burton16jmca}, characterized by the interlayer stacking called 2H-\ce{SnS2} polytype as recently studied with first-princples method~\cite{Kumagai16pra,Whittles16cm,Skelton17jpcc,Skelton17pccp}.
For modeling of the \ce{SnS2} layer, we used a $AA$-stacked \ce{SnS2} bilayer, employing the hexagonal ($2\times2$) cells in plane, which contain 4 Sn atoms and 8 S atoms in one layer.
For the composite systems of \ce{SnS2} layer with graphene or GO layer, we built sandwiched supercells, where the \ce{SnS2} bilayer is placed in between the graphene or GO layers, giving the hybrid \ce{SnS2}/G or \ce{SnS2}/GO composite.
Again, the hexagonal ($2\times2$) cells were used for modeling of \ce{SnS2} bilayer in these composite systems, while the $AA$-stacked hexagonal ($3\times3$) cells were employed for the graphene or GO layer that contains 18 C atoms or 18 C atoms and 9 O atoms in one layer.
It should be noted that the GO layer with the C/O atomic ratio of 2:1 was formed by decorating epoxy groups ($-$O$-$) on outer side of the graphene sheet.
For these supercell models, fixed cell lengths of $a=b=7.79$ \AA~and $c=30$ \AA~were used throughout the work, providing the vacuum layer thickness over 15 \AA~(see Figure S1).

All the calculations in this work were carried out using the pseudopotential plane-wave method as implemented in {\footnotesize QUANTUM ESPRESSO} package (version 6.2)~\cite{QE}.
We utilized the projector augmented wave (PAW) method to describe the interaction between ions and valence electrons~\footnote{We used C.pbesol-n-kjpaw\_psl.0.1.UPF, O.pbesol-n-kjpaw\_psl.0.1.UPF, Sn.pbesol-dn-kjpaw\_psl.0.2.UPF, S.pbesol-n-kjpaw\_psl.0.1.UPF, and Na.pbesol-spn-kjpaw\_psl.0.2.UPF, which are provided in the package.}.
The PBEsol functional~\cite{PBEsol} within the generalized gradient approximation (GGA) was used for the exchange-correlation interaction between valence electrons.
To take into account the vdW interaction between the layers, the vdW energy provided by the exchange-hole dipole moment (XDM) method~\cite{xdm} was added to the DFT total energy.
As the major computational parameters, plane-wave cutoff energies were set to be 40 Ry for wave function and 400 Ry for electron density, and the Monkhorst-Pack special $k$-points were set to be ($4\times4\times1$) for all the supersell models, providing a total energy accuracy of 5 meV per atom.
Self-consistent convergence threshold for total energy was $10^{-9}$ Ry, and the convergence threshold for atomic force in structural relaxations was $4\times10^{-4}$ Ry$\cdot$Bohr$^{-1}$. Fermi-Dirac function with a gaussian spreading factor of 0.01 Ry was applied to the Brillouin-zone integration.

To calculate the activation barriers for sodium migration, we applied the climbing image nudged elastic band (NEB) method~\cite{NEB}. During the NEB runs, the supercell dimensions were fixed, while all the atomic positions were allowed to relax until the atomic forces converge within 0.05 eV$\cdot$\AA$^{-1}$. The number of NEB images was tuned so that the distance between neighbouring NEB images was less than 1 \AA. We used the image energies and their derivatives to make an interpolation of the path energy profile that passes through every image points, as implemented in the neb.x code of {\footnotesize QUANTUM ESPRESSO} package~\cite{QE}.

\begin{figure*}[!t]
\scriptsize
\begin{center}
\includegraphics[clip=true, scale=0.5]{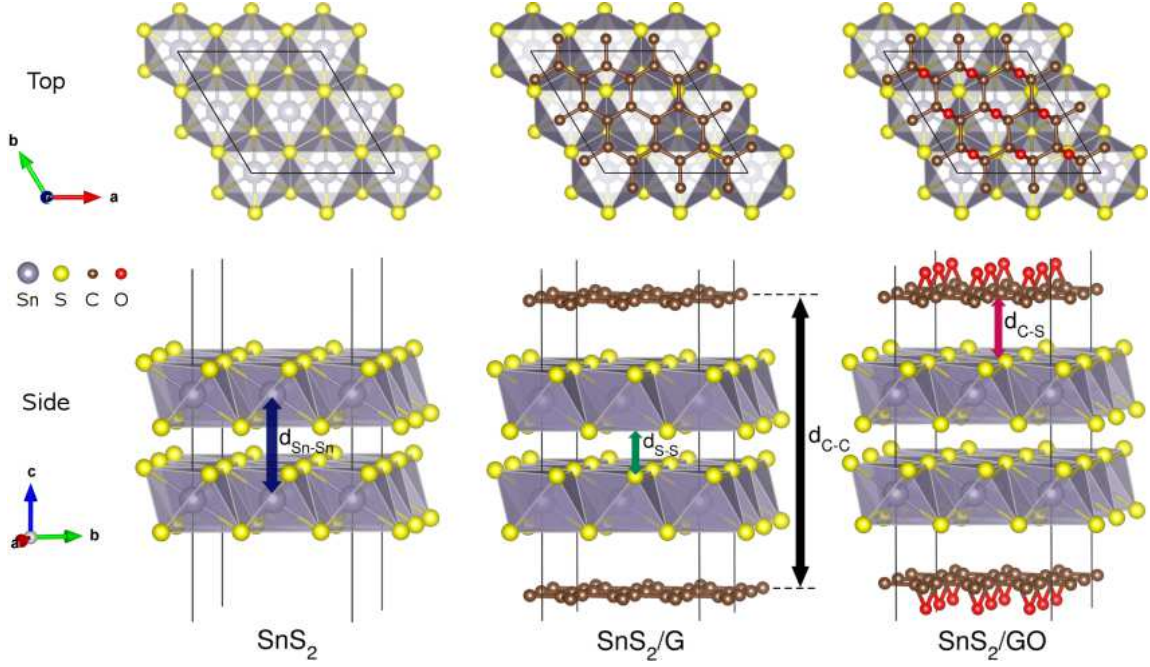}
\end{center}
\caption{\label{fig_str}Polyhedral view of relaxed atomic structures of \ce{SnS2} bilayer and hybrid \ce{SnS2}/G and \ce{SnS2}/GO composites. The hybrid composites contain the two equivalent hetero-interfaces between the \ce{SnS2} and graphene or GO layer. Various interlayer distances are denoted.}
\end{figure*}

In order to estimate the binding strength between the layers, we calculated the interlayer binding energy per atom as follows~\cite{yucj14,yucj20},
\begin{equation}
E_{\text{b}}=\frac{1}{N}[E(d=d_{\text{eq}})-E(d=d_{\infty})] \label{eq_laybind}
\end{equation}
where $E(d=d_{\text{eq}})$ and $E(d=d_{\infty})$ are the DFT total energies of the supercells with the interlayer distance at equilibrium ($d_{\text{eq}}$) and infinity ($d_{\infty}$), respectively, and $N$ is the number of atoms included in the supercell. Due to the numerical limitation, $d_{\infty}$ was approximated to be $\sim$15 \AA, over which the total energy of the supercell is scarcely changed and the interaction between the layers becomes negligible.

The possibility of interface formation can be estimated by calculating the formation energy per unit area defined as, 
\begin{equation}
E_{\text{f}}=\frac{1}{2A}[E_{\text{\ce{SnS2}/G (GO)}} - (N_{\text{\ce{SnS2}}}E^{\text{bulk}}_{\text{\ce{SnS2}}} + N_{\text{G (GO)}}E_{\text{G (GO)}})] \label{eq_form}
\end{equation}
where $E_{\text{\ce{SnS2}/G (GO)}}$ and $E_{\text{G (GO)}}$ are the total energies of the supercells including the hybrid \ce{SnS2}/G (GO) bilayer and graphene (GO) monolayer, and $E^{\text{bulk}}_{\text{\ce{SnS2}}}$ is the total energy per formula unit of unit cell of the crystalline \ce{SnS2} bulk.
$N_{\text{\ce{SnS2}}}$ and $N_{\text{G (GO)}}$ are the numbers of \ce{SnS2} formula units included in the layer and of the graphene (GO) monolayers, and $A$ is the interfacial area in $x$-$y$ plane.
Since the supercell models include two equivalent interfaces between the \ce{SnS2} and graphene or GO layers as can be seen in Figure~\ref{fig_str}, the energy difference should be divided by 2 to give the formation energy of the interface.  

To estimate the stability of sodium-intercalated compounds, we calculated the sodium intercalation energy $E_{\text{int}}$ as follows,
\begin{equation}
E_{\text{int}}=E_{\text{sub+Na}}-(E_{\text{sub}}+E^{\text{bcc}}_{\text{Na}})
\end{equation}
where $E_{\text{sub+Na}}$ and $E_{\text{sub}}$ are the total energies of the supercells for the sodium intercalated compounds and the pristine substrates (\ce{SnS2}, \ce{SnS2}/G and \ce{SnS2}/GO), and $E^{\text{bcc}}_{\text{Na}}$ is the total energy per atom of the unit cell of the sodium crystal in body-centered cubic (bcc) phase.

\section{Results and Discussion}
\subsection{Hybrid composites}
We first optimized the crystalline structure of \ce{SnS2} bulk in hexagonal phase ($P\bar{3}m1$ space group, see Figure S1) by use of different XC functionals.
In accordance with the previous first-principles calculations~\cite{Kumagai16pra,Whittles16cm,Skelton17jpcc,Skelton17pccp}, the PBEsol functional yielded slightly overestimated lattice constants (0.05\% and 2\% for $a$ and $c$) while adding the vdW energy in the flavor of Grimme ($-$0.05\% and $-$5\%) or XDM~\cite{xdm} ($-$0.7\% and $-$5\%) gave underestimation, when compared with the experimental values of $a=3.649$ \AA~and $c=5.899$ \AA~\cite{Burton16jmca}.
Although XDM gave somewhat severe underestimated interlayer lattice constant for \ce{SnS2} bulk, we choose it for further calculations of interface in this work, based on the established fact that XDM is especially good for modeling of surfaces~\cite{Hong16pccp} and organic/inorganic interfaces~\cite{Johnson17book}.

Then, we performed atomic relaxations in the supercells of the individual \ce{SnS2}, graphene and GO monolayers, of which 2D $x$-$y$  planes include $(2\times 2)$ cells for \ce{SnS2} layer and $(3\times 3)$ cells for graphene and GO layers (see Figure S1).
The cell parameters of these supercells were set to be identical and fixed during the atomic relaxations as $a=b=7.79$ \AA~and $c=30$ \AA, allowing to make a modeling of the hetero-interfaces from these individual layers.
In fact, we carried out the variable cell relaxation in only 2D $x$-$y$ plane of the GO supercell to determine the optimal cell size in the plane.
In the graphene and GO layers, the average C$-$C and C$-$O bond lengths were determined to be 1.49 and 1.42 \AA, which are comparable with the previous values of 1.51 and 1.42 \AA~obtained by first-principles calculations~\cite{Sljivancanin13c,Yan10prb,Wang10prb}.

\begin{figure}[!t]
\begin{center}
\includegraphics[clip=true,scale=0.15]{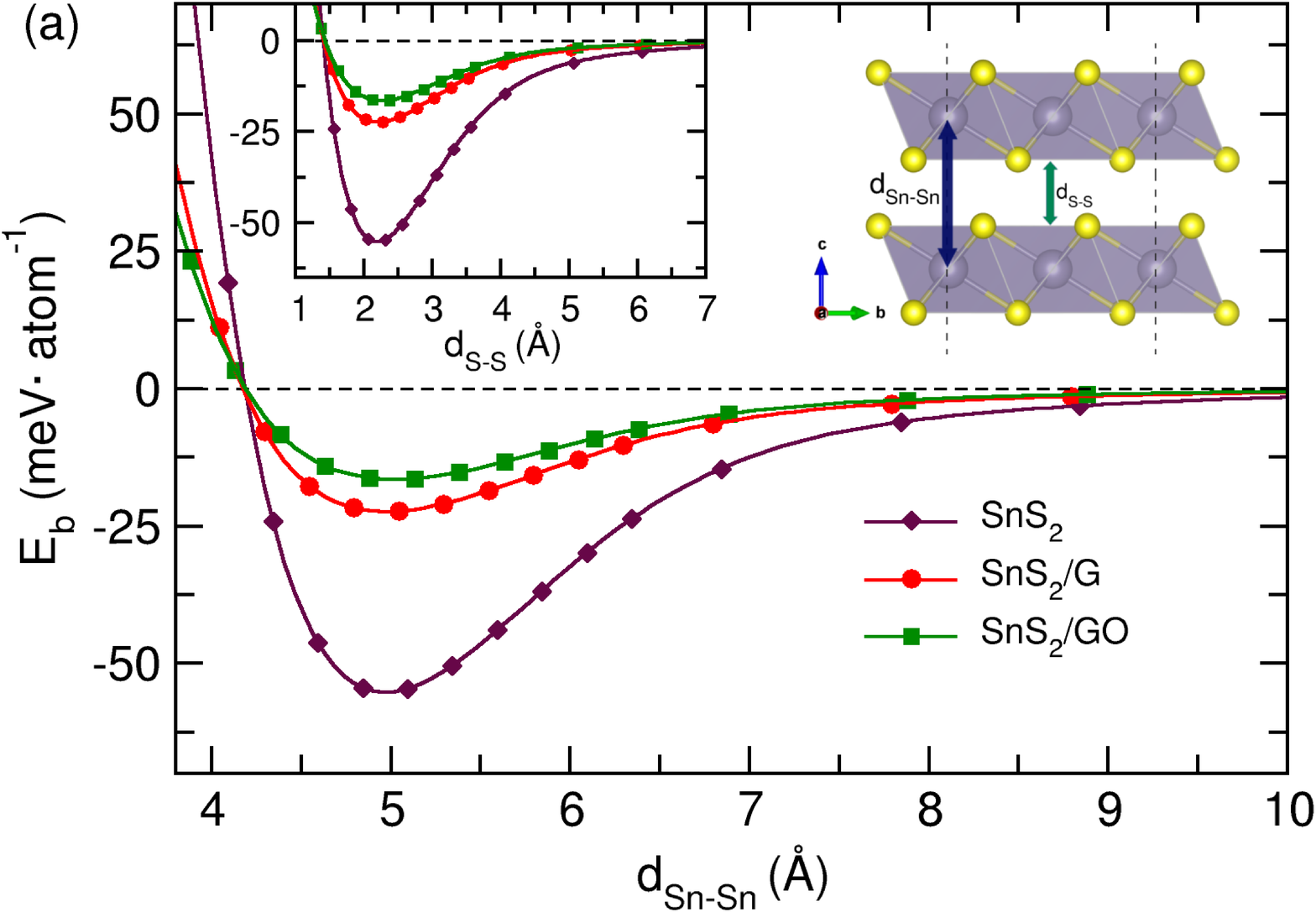} \\ \vspace{10pt}
\includegraphics[clip=true,scale=0.15]{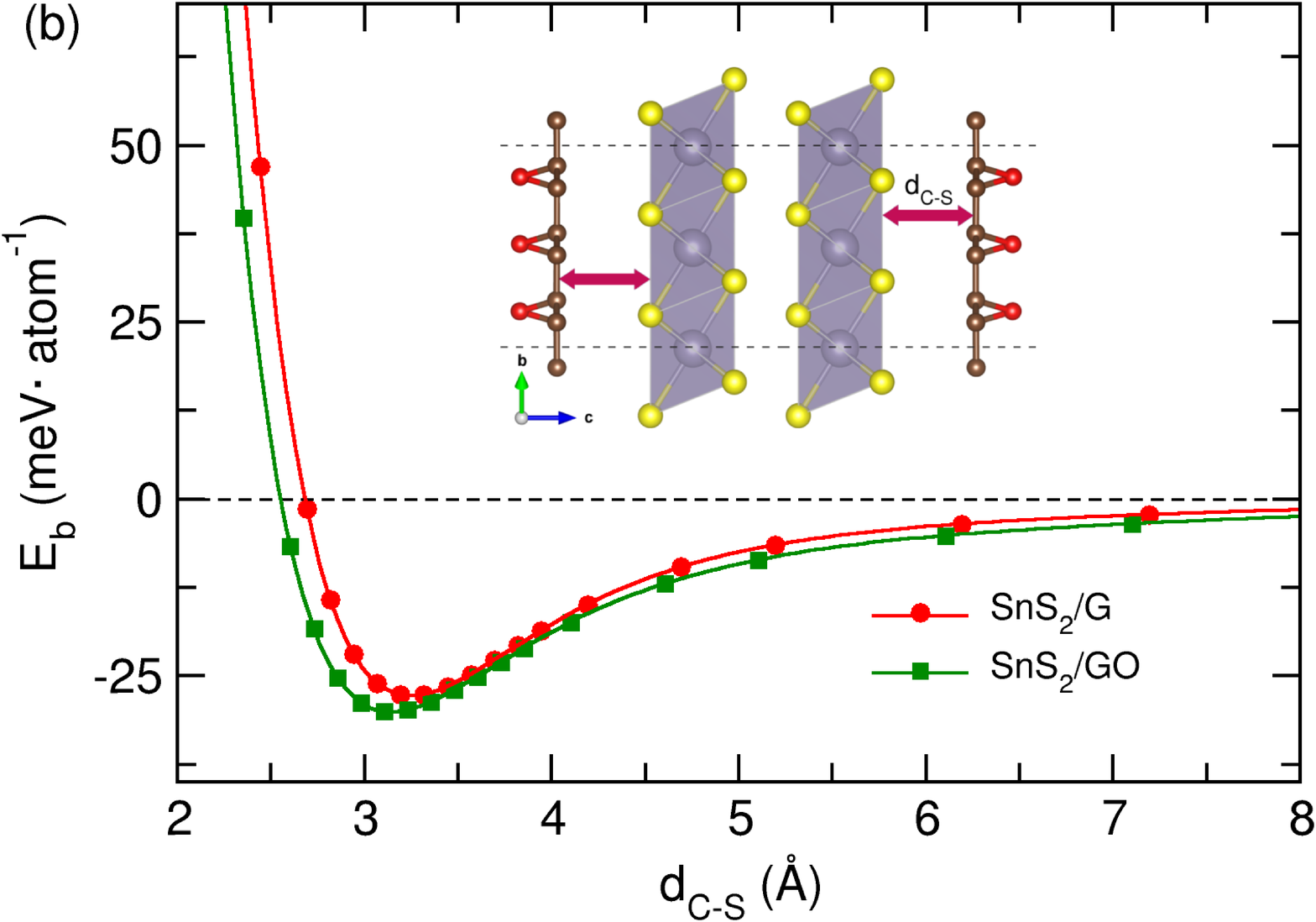}
\end{center}
\caption{\label{fig_bind}The total energy differences as varying the interlayer distances of (a) $d_{\text{Sn}-\text{Sn}}$ and $d_{\text{S}-\text{S}}$ (inset) in \ce{SnS2}, \ce{SnS2}/G and \ce{SnS2}/GO bilayers and (b) $d_{\text{C}-\text{S}}$ in \ce{SnS2}/G and \ce{SnS2}/GO bilayers.}
\end{figure}

Using these monolayers of \ce{SnS2}, graphene and GO, we made the supercells for \ce{SnS2} bilayer and hybrid \ce{SnS2}/G and \ce{SnS2}/GO composites with the same sizes to the monolayer supercells, and performed the atomic relaxations to determine the real atomic positions as shown in Figure~\ref{fig_str}.
In these bilayers, the average Sn$-$S bond lengths were found to be 2.63 \AA, which is a little larger than that of 2.56 \AA~in the bulk, while C$-$C and C$-$O bond lengths were almost same to the monolayers due to the same sizes of the supercells.

In order to determine the optimal interlayer distance and the corresponding binding energy between the layers, we estimated the total energy differences of the supercells as varying the interlayer distance.
In accordance with the variation of interlayer distance, the supercell length $c$ was also varied to preserve the vacuum layer thickness as 15 \AA.
Two different interlayer bindings can be distinguished as those between the neighboring \ce{SnS2} layers (homo-interface) and between the \ce{SnS2} and graphene or GO layers (hetero-interface), while four different interlayer distances can be thought as $d_{\text{Sn}-\text{Sn}}$, $d_{\text{S}-\text{S}}$, $d_{\text{C}-\text{S}}$ and $d_{\text{C}-\text{C}}$ as depicted in Figure~\ref{fig_str}.
Over the interlayer distance of $\sim$15 \AA, total energies of the supercells were confirmed to be almost unchangeable, indicating a non-interaction or no binding between the layers.
Figure~\ref{fig_bind} shows the total energy differences of supercells as functions of interlayer distance in reference to the total energy of supercell with the interlayer distance of $\sim$15 \AA.

\begin{table}[!t]
\caption{\label{table_nonint}Interlayer distance ($d$), interlayer binding energy per atom ($E_{\text{b}}$) and interface formation energy ($E_{\text{f}}$) in \ce{SnS2}, \ce{SnS2}/G and \ce{SnS2}/GO bilayers.}
\footnotesize
\begin{tabular}{lccc}
\hline
     & \ce{SnS2} & \ce{SnS2}/G & \ce{SnS2}/GO \\
\hline 
$d_{\text{Sn}-\text{Sn}}$ (\AA)  & 4.97 &  5.04 &  5.13 \\
$d_{\text{S}-\text{S}}$ (\AA)     & 2.18 &  2.28 &  2.37 \\
$E_{\text{b}}$ (\ce{SnS2}$-$\ce{SnS2}) (meV) & $-$54 &  $-$22 &  $-$16 \\
$d_{\text{C}-\text{S}}$ (\AA)    &          &  3.19 &  3.10 \\
$E_{\text{b}}$ (C$-$\ce{SnS2}) (meV) &        &  $-$28 &  $-$30 \\
$d_{\text{C}-\text{C}}$ (\AA)   &         & 13.96 & 13.87 \\
$E_{\text{f}}$ (J/m$^2$)  & 0.81 & 0.34 & 0.13 \\ 
\hline
\end{tabular}
\end{table}
Table~\ref{table_nonint} lists the determined optimal interlayer distances and binding energies, and interface formation energies.
The Sn$-$Sn and S$-$S interlayer distances were found to become longer upon the formation of hybrid interfaces from 4.97 and 2.18 \AA~in \ce{SnS2} bilayer to 5.04 and 2.28 \AA~in \ce{SnS2}/G, and to 5.13 and 2.37 \AA~in \ce{SnS2}/GO.
This indicates that there is an attractive interaction between \ce{SnS2} layer and graphene or GO layer, and the strength of \ce{SnS2}$-$GO interaction can be said to be higher than that of \ce{SnS2}$-$G interaction due to further lengthening of $d_{\text{Sn}-\text{Sn}}$ or $d_{\text{S}-\text{S}}$ in the case of GO hybrid.
Accordingly, the corresponding interlayer binding energy per atom between the neighboring \ce{SnS2} layers, $E_{\text{b}}$ (\ce{SnS2}$-$\ce{SnS2}), also decreases in magnitude from $-$54 meV in \ce{SnS2} to $-$22 meV in \ce{SnS2}/G, and to $-$16 meV in \ce{SnS2}/GO.
In the cases of hybrid composites of \ce{SnS2}/G and \ce{SnS2}/GO, the interlayer distance $d_{\text{C}-\text{S}}$ in the GO hybrid (3.10 \AA) was determined to be shorter than that in the graphene hybrid (3.19 \AA), while the binding energy of \ce{SnS2} layer with the GO layer ($-$30 meV) was found to be higher in magnitude than that with the graphene layer ($-$28 meV).
These also indicate that oxidation of graphene can enhance the interaction between the \ce{SnS2} layer and the graphene layer.
It is worth noting that all the calculated interlayer binding energies are smaller than the binding energy between the graphene layers in graphite ($-$77 meV by PBE+vdW calculation~\cite{yucj14,yucj20}), and all the negative values indicate certain formations of stable interface systems from the individual monolayers. 

The interlayer distances between the graphene layers ($d_{\text{C}-\text{C}}$) in the hybrid \ce{SnS2}/G and \ce{SnS2}/GO composites were calculated to be 13.96 and 13.87 \AA, indicating that the epoxy groups of GO leads to a contraction of the total interlayer distance.
The calculated formation energies of the interfaces from the \ce{SnS2} bulk and graphene or GO monolayer were turned out to be positive, suggesting an endothermic feature of the interface formations.
And it can be said from the magnitudes of their formation energies that the hetero-interface in \ce{SnS2}/G or \ce{SnS2}/GO is easier to form than the homo-interface in \ce{SnS2}, and of the two hetero-interfaces the \ce{SnS2}/GO is easier than the \ce{SnS2}/G.

\subsection{Sodium intercalation and migration}
\begin{figure}[!t]
\scriptsize
\begin{center}
\includegraphics[clip=true,scale=0.18]{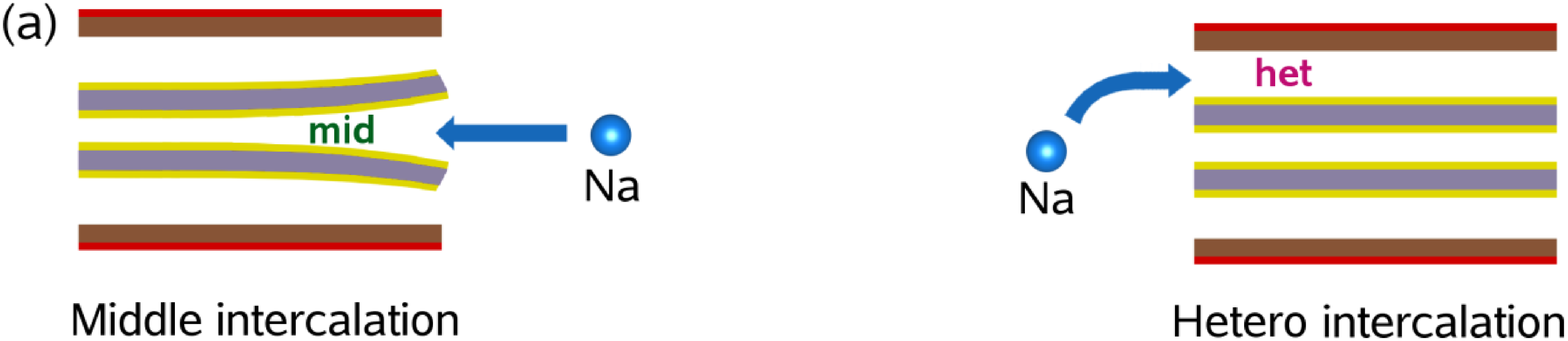} \\ \vspace{10pt}
\includegraphics[clip=true,scale=0.18]{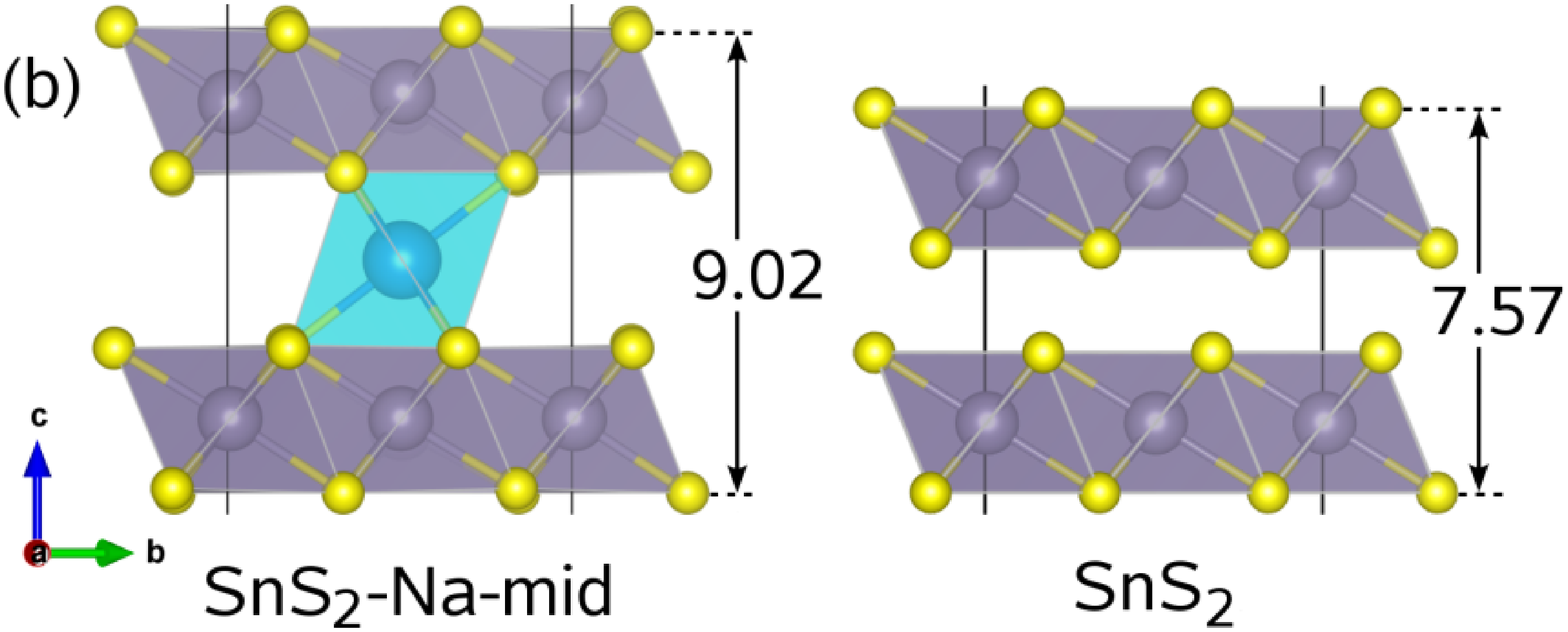} \\ \vspace{10pt}
\includegraphics[clip=true,scale=0.18]{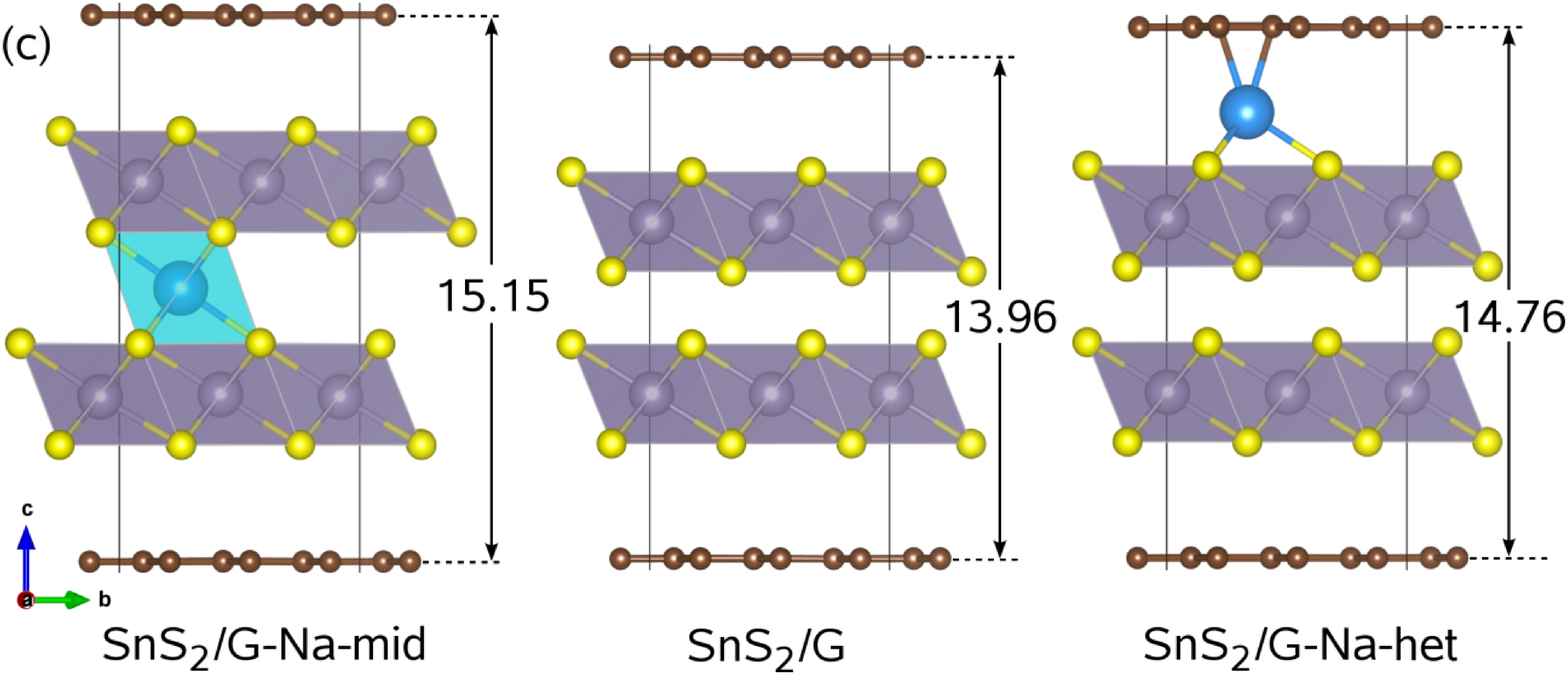} \\ \vspace{10pt}
\includegraphics[clip=true,scale=0.18]{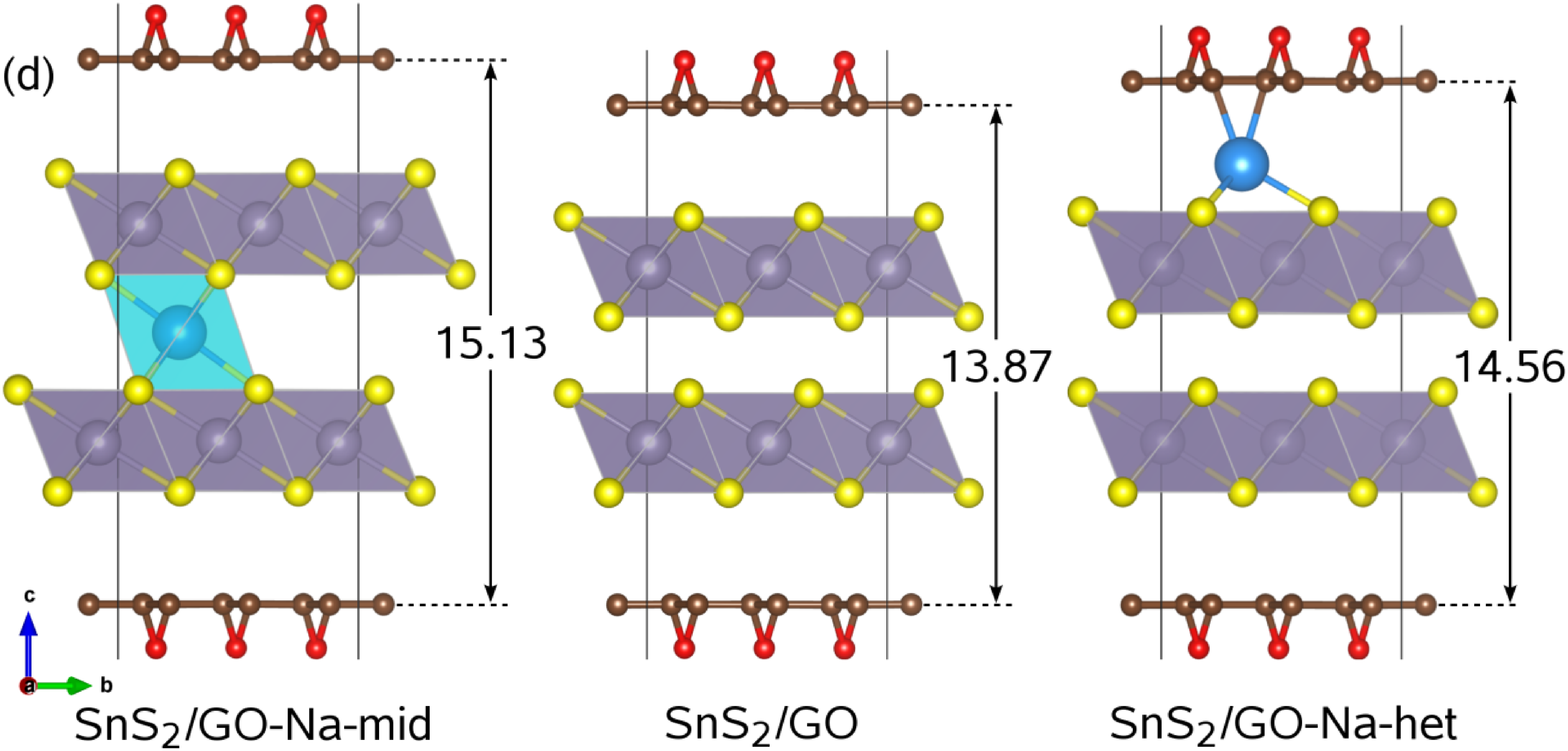}
\end{center}
\caption{\label{fig_intstr}(a) Schematic illustration of sodium intercalation into the middle and hetero interspaces of \ce{SnS2} and graphene or GO composites. Side views of relaxed atomic structures of sodium intercalated (b) \ce{SnS2}, (c) \ce{SnS2}/G, and (d) \ce{SnS2}/GO compounds. Left panel is for the Na middle intercalations, central panel for the substrates, and right panel for the Na hetero intercalations. Typical interlayer distances in \AA~unit are denoted for comparison.}
\end{figure}

To study the sodium intercalation into these \ce{SnS2} composites containing the interfaces, we considered two different scenarios: (1) sodium intercalation into the middle interspace between the neighboring \ce{SnS2} layers, forming {\it substrate-Na-mid} compounds, and (2) sodium insertion into the hetero interspace between the \ce{SnS2} and graphene or GO layers, forming {\it substrate-Na-het} compounds, as illustrated in Figure~\ref{fig_intstr}(a).
For the case of \ce{SnS2} bilayer, only the middle intercalation is surely possible.

In each way of sodium intercalation, we identified the energetically favorable anchoring site of sodium atom by comparing the total energies of the supercells with different sodium insertion sites.
In the cases of middle intercalation, two different sites might be thought: (1) Na atom can sit on the top of S atom of the below \ce{SnS2} layer while at the center of S-triangle of the upper \ce{SnS2} layer, leading to a formation of \ce{NaS4} tetrahedron ({\it tet} site), and (3) Na atom can be placed at the spherical center of \ce{SnS6} octahedron ({\it oct} site).
In energetics the Na middle-intercalated \ce{SnS2} supercell in the {\it tet} site was found to be about 0.39 eV higher than that in the {\it oct} site, possibly due to a stronger bonding of Na atom with the surrounding S atoms (see Figure S2).
Henceforth, only the {\it oct} site will be considered for the middle intercalations.
Furthermore, it was observed for the cases of hybrid \ce{SnS2}/G (GO) composites that the \ce{SnS2} layers can be shifted in opposite way each other, resulting in the change of layer stacking from $AA$ to $AB$ with a lowering of total energy as 0.21 eV (0.17 eV) (see Figure S2).
On the other hand, we distinguished three different sites for the sodium hetero-intercalations in the cases of \ce{SnS2}/G (GO) composites: (1) Na atom can sit on the top of Sn atom placed at the center of S-triangle in the below \ce{SnS2} layer while on the side of C$-$C bond of the upper graphene (GO) layer ({\it top} site), (2) on the hollow site at the center of S-triangle in the below layer ({\it hollow} site), and (3) at the center of carbon hexagon ring of the upper layer ({\it center} site). Among these sodium hetero-intercalation sites, the {\it top} site was found to have the lowest total energy (see Figure S3). 

\begin{table}[!b]
\footnotesize
\caption{\label{table_int}Overview of sodium intercalation and diffusion in the both middle and hetero ways, including the interlayer distances $d$, the volume expantion ratio $r$, the sodium intercalation energy $E_\text{int}$ and the sodium migration energy $E_\text{mig}$.}
\begin{tabular}{lc@{}ccc@{}ccc}
\hline
  &\ce{SnS2} & & \multicolumn{2}{c}{\ce{SnS2}/G} & & \multicolumn{2}{c}{\ce{SnS2}/GO} \\ 
\cline{2-2} \cline{4-5} \cline{7-8}
    & mid & & mid & het & & mid & het \\
\hline
$d_{\text{Sn}-\text{Sn}}$ (\AA) & 6.21 & & 5.91 & 4.86 & & 6.02 & 4.93 \\
$d_{\text{S}-\text{S}}$ (\AA) & 3.37 & & 3.09 & 2.10 & & 3.19 & 2.17 \\
$d_{\text{C}-\text{S}}$ (\AA) &  & & 3.26 & 3.87 & & 3.19 & 3.65 \\
$d_{\text{C}-\text{C}}$ (\AA) &  & & 15.15 & 14.76 & & 15.13 & 14.56 \\
$r$ (\%)                     &   119   & & 109   & 106   & & 109  & 105 \\
$E_{\text{int}}$ (eV) & $-$0.30 & & $-$1.29 & $-$0.76 & & $-$1.12 & $-$0.35 \\
$E_{\text{mig}}$ (eV) & 0.52 & & 0.37 & 0.32 & & 0.48 & 0.38 \\  
\hline
\end{tabular}
\end{table}
Figure~\ref{fig_intstr} shows the relaxed atomic structures of sodium-intercalated \ce{SnS2} and \ce{SnS2}/G (GO) composites in the lowest energy configurations.
It is clear that the volume expansion ratio between the Na-intercalated composite ($V_{\text{int}}$) and the substrate ($V_0$), $r=\frac{V_{\text{int}}}{V_0}\times 100$\%, which can be readily estimated by using the outermost interlayer distances due to the same in-plane surface area, is distinctly larger in the  non-hybrid \ce{SnS2} bilayer ($\sim$119\%) than in the hybrid \ce{SnS2}/G (GO) composites.
This indicates that the hybrid of \ce{SnS2} with graphene or GO can effectively reduce the big volume change of \ce{SnS2} itself upon de-/sodiation process.
For the both cases of hybrid \ce{SnS2}/G and \ce{SnS2}/GO composites, meanwhile, the volume expansion ratios upon Na middle intercalations (109\%) are larger than the hetero intercalations (105\%).
As presented in Table~\ref{table_int}, the interlayer distances such as $d_{\text{Sn}-\text{Sn}}$, $d_{\text{S}-\text{S}}$ and $d_{\text{C}-\text{C}}$ are clearly smaller in the Na hetero-intercalated compounds than in the middle-intercalated ones.
From these results, the sodium intercalation into the hetero-interface between the \ce{SnS2} layer and graphene or GO layer is more profitable for SIB application than into the homo-interface with respect to the volume change.
The intercalation energies were calculated to be negative, indicating an exothermic reaction of sodium intercalation from the bilayer substrate and sodium metal.
The magnitude of the intercalation energy for the cases of hetero-interfaces was smaller than for the homo-interface, giving an expectation of good performance of Na hetero-intercalated \ce{SnS2}/G (GO) as anode materials, which need to possess an electrode potential as low as possible.

\begin{figure}[!t]
\begin{center}
\begin{tabular}{c}
\includegraphics[clip=true,scale=0.32]{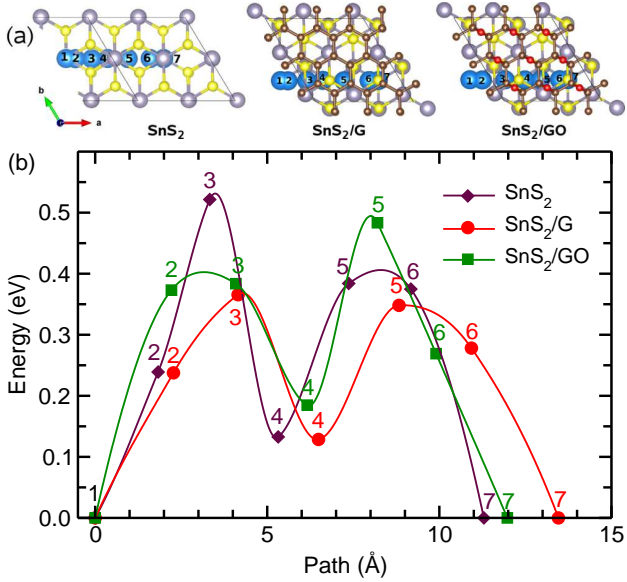} \\
\includegraphics[clip=true,scale=0.5]{fig4b.eps}
\end{tabular}
\end{center}
\caption{\label{fig_migmid}(a) Top view of sodium migrations along the 1D lines within the middle interspaces of \ce{SnS2}, \ce{SnS2}/G and \ce{SnS2}/GO composites, and (b) the corresponding path energy profiles.}
\end{figure}
We then investigated the diffusion of sodium atom by calculating their activation barriers, which can give a suggestion of the rate capability and cycling stability of electrode materials in SIB application.
For the cases of middle intercalation, we considered the sodium migration path from the site determined by atomic relaxation to the adjacent image site, which consists of 7 NEB image configurations.
It should be noted that the interval of the neighboring image points was set to be around 1 \AA, and the migration paths were determined to be almost straight line along the $a$ direction on the $x$-$y$ plane, as shown in Figure~\ref{fig_migmid}(a).
In Figure~\ref{fig_migmid}(b), we show the path energy profile corresponding to the migration paths, from which the activation energy barriers were determined to be 0.52, 0.37 and 0.48 eV for the case of \ce{SnS2}, \ce{SnS2}/G and \ce{SnS2}/GO, respectively.
Due to the lower activation energies in the hybrid \ce{SnS2}/G or GO composites than in the pristine \ce{SnS2} itself, it can be said that the hybrid of \ce{SnS2} with graphene or GO layer can facilitate the diffusion of sodium. 

\begin{figure}[!t]
\begin{center}
\begin{tabular}{l}
~\includegraphics[clip=true,scale=0.3]{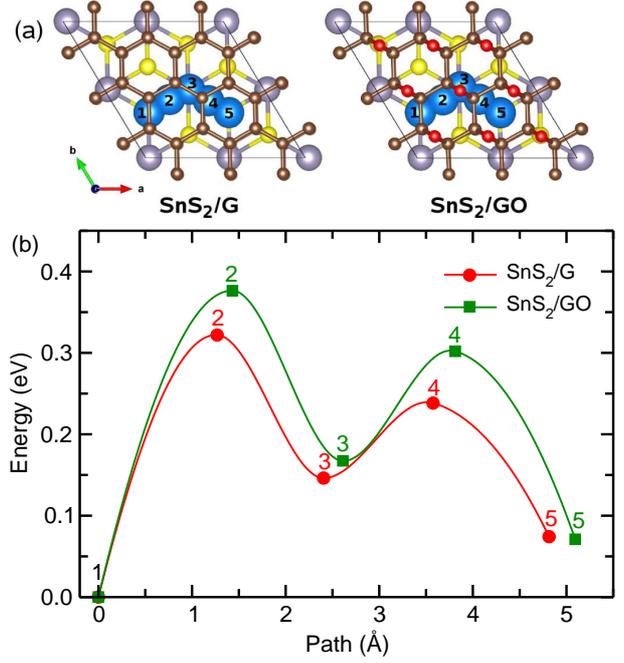} \\
\includegraphics[clip=true,scale=0.5]{fig5b.eps}
\end{tabular}
\end{center}
\caption{\label{fig_mighetero}(a) Top view of sodium migrations within the hetero interspaces of hybrid \ce{SnS2}/G and \ce{SnS2}/GO composites, and (b) the corresponding path energy profiles.}
\end{figure}
Figure~\ref{fig_mighetero} presents the results of the sodium migration within the hetero-interface between the \ce{SnS2} and graphene or GO layer.
We used 5 NEB image congfigurations due to the shorter length of path.
As shown in Figure~\ref{fig_mighetero}(a), the migration paths are 2D on the $x$-$y$ plane, passing the {\it top}, {\it hollow} and {\it center} sites, as described above subsection.
In Figure~\ref{fig_mighetero}(b), we show the path energy profiles in the hybrid \ce{SnS2}/G and \ce{SnS2}/GO composites, and the activation barriers as 0.32 and 0.38 eV, which are clearly lower than the corresponding sodium migrations within the middle interface.
The migration barriers are also listed in Table~\ref{table_int}.
To sum up, the sodium atom can migrate more smoothly in the hetero-interspace than in the middle-interspace, and hybrid with graphene is slightly more beneficial than with graphene oxide.
It is worth noting that the calculated activation energies are comparable with those for sodium or lithium migration in graphite (0.40 eV)~\cite{Nobuhara13jps} and for sodium-molecule migration (0.40 eV)~\cite{yucj14}.
      
\subsection{Electronic properties}
To further understand such enhancement of sodium insertion by hybrid of \ce{SnS2} with graphene or GO and make it clear the role of hetero-interface, we carried out the analysis of electronic properties such as density of state (DOS) and electronic charge density differneces.
Such electronic properties can provide us valuable insight for the interaction between the layers and charge transfer in the event of hetero-interface formation, which can be useful for understanding the enhancement of electrochemical properties in the 2D hybrid materials. 
To obtain an intuitive insight for charge transfer between \ce{SnS2} and graphene or GO layers during the interface formation, we calculated the electronic charge density difference ($\Delta\rho$) as follows,
\begin{equation}
\Delta\rho = \rho_{\ce{SnS2}/\text{G (GO)}}-(\rho_{\ce{SnS2}}+\rho_{\text{G (GO)}}) 
\end{equation}
where $\rho_{\ce{SnS2}/\text{G (GO)}}$, $\rho_{\ce{SnS2}}$ and $\rho_{\text{G (GO)}}$ are the charge densities of the hybrid composites, individual \ce{SnS2} and graphene or GO layers, respectively.

\begin{figure}[!t]
\scriptsize
\begin{center}
\includegraphics[clip=true,scale=0.23]{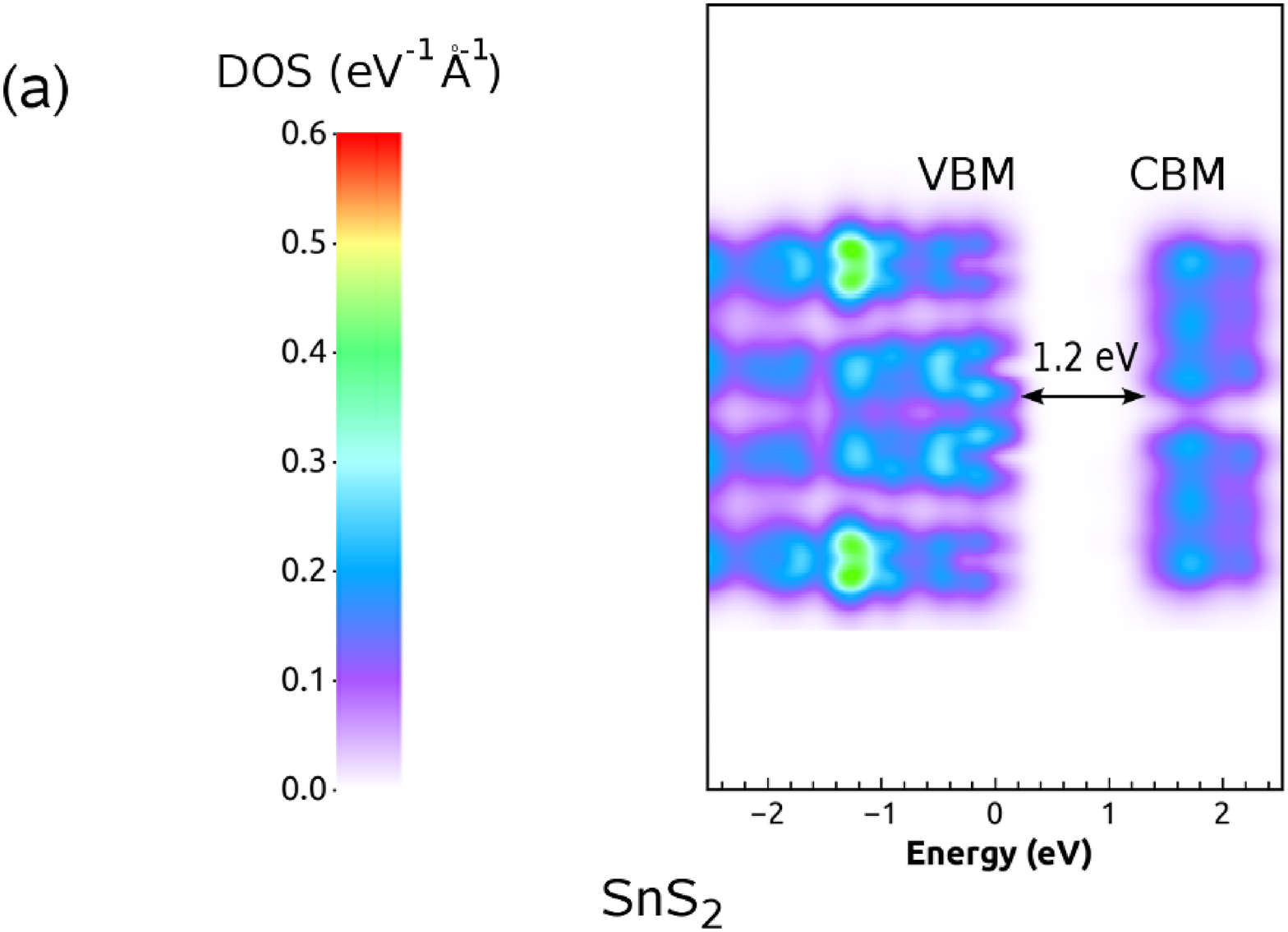} \\ \vspace{5pt}
\includegraphics[clip=true,scale=0.23]{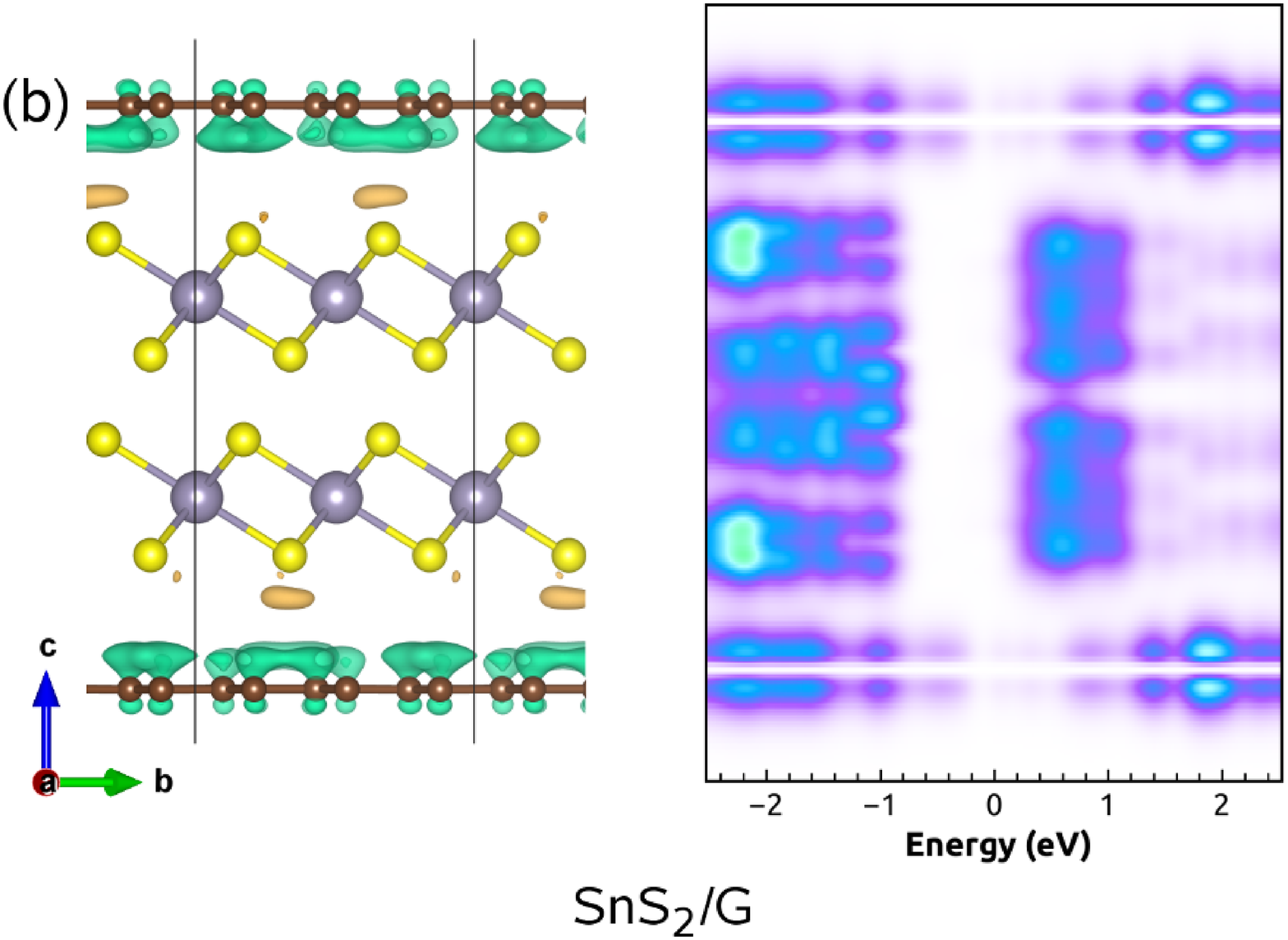} \\ \vspace{5pt}
\includegraphics[clip=true,scale=0.23]{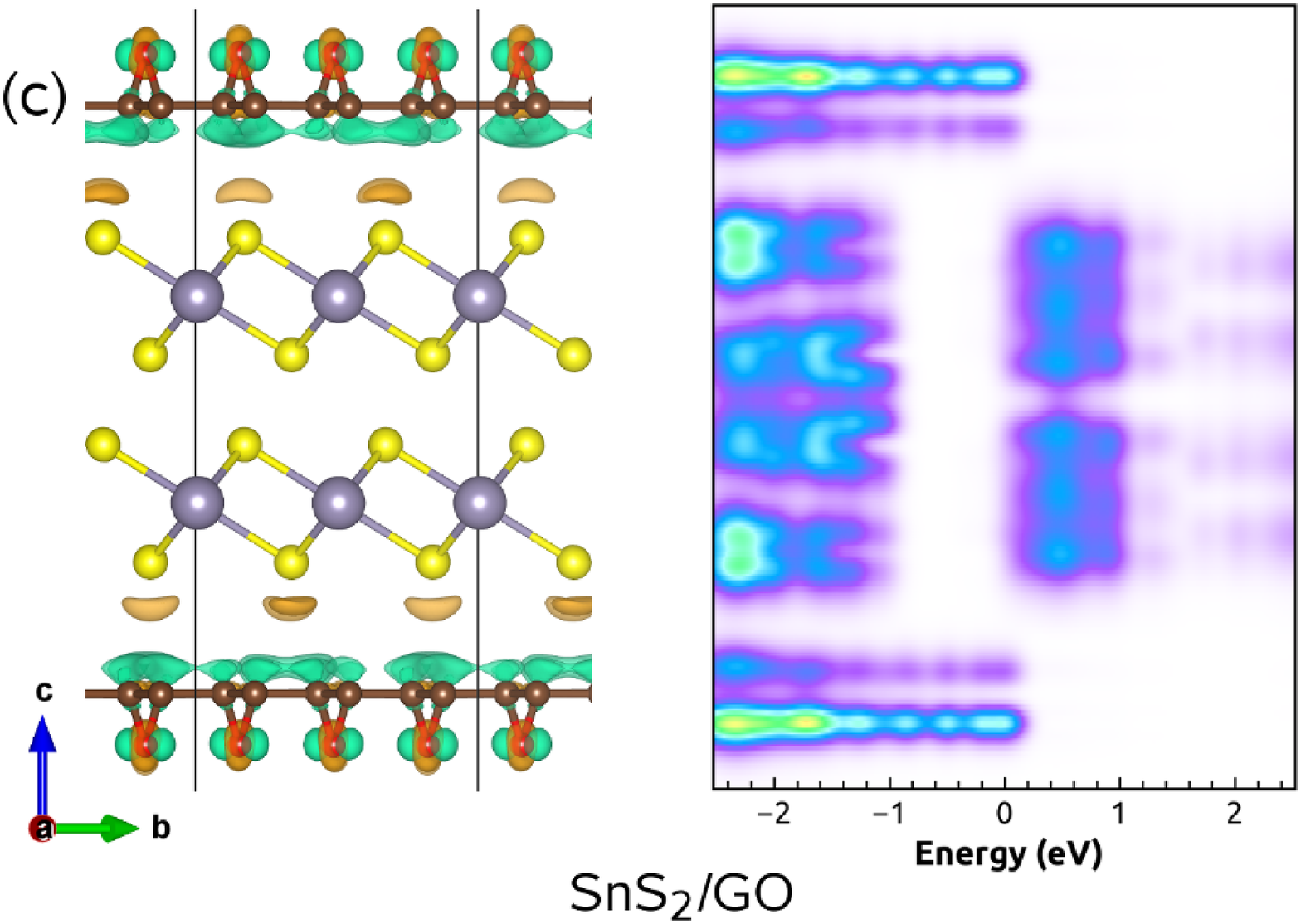}
\end{center}
\caption{\label{fig_chargedos}Isosurface plot of electronic charge density differences (left panel) and integrated local density of state (right panel) in (a) \ce{SnS2} bilayer, (b) \ce{SnS2}/G and (c) \ce{SnS2}/GO composites. The isosurface values are $\pm$0.0005 $|e|\cdot$\AA$^{-3}$ in (b) \ce{SnS2}/G and $\pm$0.0013 $|e|\cdot$\AA$^{-3}$ in (c) \ce{SnS2}/GO. Brown (green) color represents the electron accumulation (loss). The Fermi level is set to be zero.}
\end{figure}
In Figure~\ref{fig_chargedos}, we show the isosurface plot of the electronic charge density differences in the hybrid \ce{SnS2}/G and \ce{SnS2}/GO composites at the value of $\pm$0.0005 and $\pm$0.0013 $|e|\cdot$\AA$^{-3}$, respectively.
For the case of hybrid \ce{SnS2}/G composite, it is shown that the carbon atoms of graphene layer lose some electrons as represented by green-colored charge depletion, while the tin atoms gain the electrons as represented by brown-colored charge accumulation (Figure~\ref{fig_chargedos}(b) left panel).
In Figure~\ref{fig_chargedos}(c) for the hybrid \ce{SnS2}/GO composite, we can also see the electronic charge accumulation around the oxygen atoms as well as the tin atoms, while the charge depletion can be seen around the carbon atoms of GO layer as well.
In these hybrid composites, therefore, it is clear that the carbon atoms play as electron donors, whereas the tin and oxygen atoms play as electron acceptors, indicating the strong interaction between the \ce{SnS2} layer and graphene or GO layer and thus stable formation of the hetero-interfaces.
Due to the influence of oxygen atoms of epoxy groups in the GO hybrid, the extent of charge transfer from the carbon atoms to the oxygen and tin atoms seems to be enhanced compared with the graphene hybrid, leading to the slight strengthening of interlayer interaction between the GO and \ce{SnS2} layers.
When intercalating the Na atom into the bilayerd substrates, the Na and Sn atoms were found to give electrons, leading to the enhanced interaction between the layers (see Figure S4).

We present the local density of state (LDOS) integrated on $x$-$y$ plane and thus along the $z$ axis in the right panel of Figure~\ref{fig_chargedos}, where the Fermi level is set to be zero.
As shown in Figure~\ref{fig_chargedos}(a), the \ce{SnS2} bilayer model was found to have a band gap of $\sim$1.2 eV between the valence-band maximum (VBM) and the conduction-band minimum (CBM), which is smaller than that of bulk (2.28 eV)~\cite{Whittles16cm}, indicating a semiconducting feature of \ce{SnS2} bilayer.
In the right panel of Figure~\ref{fig_chargedos}(b) and (c) for the hybrid cases, however, the Fermi level can be seen to be very close to the CBM belonged to the \ce{SnS2} layer, resulting in an enhancement of electronic conductivity upon hybrid of \ce{SnS2} with graphene or GO.
Moreover, we can see the occupied valence states of graphene layer and denser LDOS of GO layer close to the Fermi level in these hybrids, indicating a good in-plane electronic conductivity due to the $p_z$ electron states as already well known before~\cite{yucj06,yucj14,yucj20}.
If the sodium atom intercalated into the hetero-interface, the \ce{SnS2} layers get fill their unoccupied states by the injected electrons from the sodium atom, and thus the Fermi level is placed on the CBM (see Figure S4).
These reveal the enhancement of electrochemical properties of the hybrid \ce{SnS2}/G (GO) composites compared to the pristine \ce{SnS2} bilayer by formation of hetero-interface through the electron trasnfer from the graphene or GO layer to the \ce{SnS2} layer.

\section{Conclusions}
In this work we provide a valuable understanding of material properties of 2D hybrid composites composed of \ce{SnS2} and graphene or GO layers that can be used as effective anode materials of sodium-ion battery.
Various supercell models of the pristine \ce{SnS2} bilayer and the hybrid \ce{SnS2}/G or GO composites have been suggested, and their atomic structures and energetics were investigated by using first-principles calculations.
Through the analysis of the calculated interlayer distances and binding and formation energies, it was revealed that the hetero-interface between the \ce{SnS2} layer and graphene or GO layer can be formed, while oxidation of graphene can enhance the interaction between the layers.
Furthermore, we have devised two different models for sodium intercalation such as middle intercalation into the interspace between the \ce{SnS2} layers and hetero intercalation into the interspace between the \ce{SnS2} and graphene or GO layers, and it was found that the hetero-intercalation is more probable due to higher intercalation energy, lower migration energy and smaller volume change.
To gain a deep understanding of material properties of these 2D composites, the electronich charge density differences and integrated local density of state were investigated, demonstrating the electron transfer from the graphene layer to the \ce{SnS2} layer and the improvement of electronic conduction by formation of hybrid composites.
In conclusion, the 2D hybrid \ce{SnS2}/G or GO composite including the hetero-interface can be a promising candidate for anode material of sodium-ion battery.

\section*{Associated Content}
\textbf{Supporting Information} \\
The Supporting Information is available free of charge on the ACS Publications website at DOI: xxx/xxx.

\section*{\label{auth}Author information}
\textbf{Corresponding Author} \\
*E-mail: cj.yu@ryongnamsan.edu.kp (Chol-Jun Yu) \\
\textbf{ORCID} \\
Chol-Jun Yu: 0000-0001-9523-4325 \\
\textbf{Notes} \\
The authors declare no competing financial interest.

\section*{\label{ack}Acknowledgments}
This work is supported as part of the fundamental research project ``Design of Innovative Functional Materials for Energy and Environmental Application'' (No. 2016-20) funded by the State Committee of Science and Technology, DPR Korea. Computation has done on the HP Blade System C7000 (HP BL460c) that is owned by Faculty of Materials Science, Kim Il Sung University.

\bibliographystyle{elsarticle-num-names}
\bibliography{Reference}

\end{document}